\pdfoutput=1
\documentclass[a4paper,fleqn,usenatbib]{mnras}

\usepackage[T1]{fontenc}
\usepackage{ae,aecompl}

\usepackage{graphicx}	
\usepackage{amsmath}	
\usepackage{amssymb}	

\newcommand{\polang}{\psi}                  
\newcommand{\DeltaAng}{\Delta \polang}          
\newcommand{\vect}[1]{\vec{#1}} 
\newcommand{\lag}{\delta}

\title[]{Matching dust emission structures and magnetic field in high-latitude cloud L1642: comparing $Herschel$ and $Planck$ maps\thanks{{\it Herschel} is an ESA space observatory with science instruments provided by European-led Principal Investigator consortia and with important participation from NASA.}}

\author[J. Malinen, L. Montier et al.]{J. Malinen,$^{1}$\thanks{E-mail:johanna.malinen@alumni.helsinki.fi} L. Montier,$^{2,3}$ J. Montillaud,$^{4}$ M. Juvela,$^{5}$ I. Ristorcelli,$^{2,3}$
\newauthor
S. E. Clark,$^{6}$ O. Bern\'e,$^{2,3}$ J.-Ph. Bernard,$^{2,3}$ V.-M. Pelkonen,$^{4}$ and D. C. Collins$^{1}$
\\
$^{1}$Department of Physics, Florida State University, Tallahassee, FL, USA\\
$^{2}$Universit\'e de Toulouse, UPS-OMP, IRAP, F-31028 Toulouse cedex 4, France\\
$^{3}$CNRS, IRAP, 9 Av. colonel Roche, BP 44346, F-31028 Toulouse cedex 4, France\\
$^{4}$Institut Utinam, CNRS UMR 6213, OSU THETA, Universit\'e de Franche-Comt\'e, 41bis avenue de l'Observatoire, 25000 Besan\c{c}on, France\\
$^{5}$University of Helsinki, P.O. Box 64, FI-00014 Helsinki, Finland\\
$^{6}$Department of Astronomy, Columbia University, New York, NY, USA
}

\date{Accepted XXX. Received YYY; in original form ZZZ}

\pubyear{2015}

\begin{document}
\label{firstpage}
\pagerange{\pageref{firstpage}--\pageref{lastpage}}
\maketitle

\begin{abstract}
The nearby cloud L1642 is one of only two known very high latitude (|b| > 30 deg) clouds actively forming stars. It is a rare example of star formation in isolated conditions, and can reveal important details of star formation in general, e.g., of the effect of magnetic fields. We compare $Herschel$ dust emission structures and magnetic field orientation revealed by $Planck$ polarization maps in L1642. The high-resolution ($\sim20\arcsec$) $Herschel$ data reveal a complex structure including a dense, compressed central clump, and low density striations. The $Planck$ polarization data (at 10$'$ resolution) reveal an ordered magnetic field pervading the cloud and aligned with the surrounding striations. There is a complex interplay between the cloud structure and large scale magnetic field. This suggests that the magnetic field is closely linked to the formation and evolution of the cloud. CO rotational emission confirms that the striations are connected with the main clumps and likely to contain material either falling into or flowing out of the clumps. There is a clear transition from aligned to perpendicular structures approximately at a column density of $N_{\rm{H}} = 1.6 \times 10^{21}\, {{\rm cm}}^{-2}$. Comparing the $Herschel$ maps with the $Planck$ polarization maps shows the close connection between the magnetic field and cloud structure even in the finest details of the cloud.
\end{abstract}




\section{Introduction} \label{sect:introduction}

The major physical processes involved in molecular cloud and star formation are gravitation, turbulence, magnetic fields, and thermal pressure, but the full picture is not clear, especially regarding the relative importance of turbulence and magnetic fields \citep[e.g.,][]{McKee2007, Bergin2007,Crutcher2012,Andre2014}. The Galactic magnetic field covers our whole Galaxy and takes part in the dynamics of the interstellar medium (ISM) and in the different phases of the star formation process, from molecular clouds to filaments and cores where stars are born.

Magnetic fields can be studied using several methods, including Zeeman splitting of spectral lines, and polarization of starlight or thermal dust emission. In this paper we use the last method, polarized dust emission. 
Because dust grains are not spherical, their radiation is polarized along the main grain axis. 
When grains are aligned with a magnetic field, the observed thermal radiation is linearly polarized~\citep[e.g.,][]{Davis1951,Vaillancourt2007}. The grains align with their long axis perpendicular to the magnetic field orientation. Therefore, if the polarization vectors are rotated by 90 degrees, the dust observations reveal the magnetic field orientation in the plane of the sky (POS).

Ground-based observations of polarized starlight in diffuse areas \citep[e.g.,][]{Myers1991,Pereyra2004} and dust emission in dense areas \citep[e.g.,][]{Ward-Thompson2000, Crutcher2004} have shown that the POS magnetic field lines are linked to the cloud structures.

Recently, $Planck$ satellite has observed the whole sky in submm-mm wavelengths, including the polarization with high enough sensitivity to map magnetic fields both in dense and diffuse areas~\citep{Planck2014I}. \citet{Planck2015XIX} presented an overview of the $Planck$ polarization data, and reported a systematic decrease of the polarization fraction with increasing column density. \citet{Planck2015XX} compared the polarization observations with magnetohydrodynamical (MHD) simulations, and came to the same conclusion that the polarization fraction is highest in the most diffuse areas. \citet{Planck2014XXXII} studied large, localized filamentary structures, which they called "ridges", in the Galaxy having hydrogen column densities ($N_{\rm{H}}$) between 10$^{20}$ and 10$^{22}$ cm$^{-2}$, and found that these structures can be seen also in the polarization data (Stokes Q and U maps). The structures are usually aligned with the magnetic field, especially at lower column densities. \citet{Planck2014XXXIII} analysed and modelled three filaments in more detail. \citet{Planck2015XXXV} made a quantitative analysis of the orientation of the magnetic field and column density structures in ten nearby clouds, finding that the relative orientation is likely to change from preferentially parallel or without a preferred orientation to perpendicular with increasing column density,
the change occuring at $N_{\rm{H}} \sim 10^{21.7} \rm{cm}^{-2}$.

Several observational studies have revealed more details in the diffuse ISM structure. \citet{Goldsmith2008} and \citet{Palmeirim2013} compared earlier magnetic field observations with CO data and $Herschel$ dust emission maps, respectively, concluding that striations in the diffuse ISM tend to be parallel to magnetic field lines and perpendicular to denser filaments. \citet{Clark2014,Clark2015} found linear, several degrees long structures, "fibers", at high Galactic latitude in neutral hydrogen (HI) maps at 4$'$ resolution. They compared HI to polarized starlight and $Planck$ dust polarization, and concluded that the "fibers" are closely aligned with magnetic fields.

As observations reveal more and more details of elongated structures at different size scales of the interstellar medium (ISM) and molecular clouds, the nomenclature is still developing and sometimes ambiguous between different subfields. Here, we use the following definitions for different types of structures, most of them elongated or "filamentary". By "filament" we mean an elongated, medium-low column density structure (approximately at $N_{\rm{H}} > 2\times 10^{21}\rm{cm}^{-2}$, but this is not a strict limit) linked to low-mass star forming regions. By "fiber" we mean elongated sub-structure of filaments~\citep[as used in][]{Andre2014}. By "striation" we mean any low column density structure (approximately at $N_{\rm{H}} < 2 \times 10^{21} \rm{cm}^{-2}$) which is either linear or curving, and either at diffuse areas or connected to denser filaments or other structures. By "blob" we mean an irregular structure of relatively dense matter which is more round than filamentary.

\citet{Malinen2014} presented $Herschel$ dust emission maps of molecular cloud L1642, which showed clear striations in the diffuse matter surrounding the cloud. L1642 is one of only two high-latitude ($|b| > 30^{\circ}$) clouds confirmed to have active star formation. 
The high galactic latitude ($-36.55^{\circ}$) with very low line-of-sight contamination, small distance (140 pc or less,~\citet{Hearty2000, Schlafly2014}), and relatively low column density make the cloud a good object for studying low-mass star formation~\citep[see][]{McGehee2008,Malinen2014}.

Our aim in this paper is to study the magnetic fields revealed by $Planck$ dust polarization maps and compare them to the striations and other structures shown in the higher resolution $Herschel$ maps of L1642. Previous $Planck$ papers have used the polarization maps only in combination with the $Planck$ intensity maps, at best at $\sim$5$'$ resolution, but usually convolved to lower resolution ($\sim$10$\arcmin$). We will show that the $Planck$ magnetic field maps can be very useful even when compared to higher resolution intensity maps, like $Herschel$ at $\sim$20$"$-40$"$ resolution. 

The contents of this article are the following: We present the observations and data processing in Sect.~\ref{sect:observations}, and methods in Sect.~\ref{sect:methods}. We analyze the cloud structure using $Herschel$ maps and compare it to the magnetic fields and kinematic data in Sect.~\ref{sect:results}. We discuss the implications of the results in Sect.~\ref{sect:discussion} and draw our conlusions in Sect.~\ref{sect:conlusions}.

\section[]{Observations and data processing} \label{sect:observations}

\subsection{Herschel}

L1642 (G210.90-36.55) was observed by $Herschel$~\citep{Pilbratt2010} as part of the Galactic Cold Cores project~\citep{Juvela2012}. The data were presented and described in~\citet{Malinen2014}. Here, we use $Herschel$ 250 $\mu$m dust emission map, which shows striations of the diffuse material surrounding the cloud, and hydrogen column density map for separating dense and diffuse areas. We have calculated the column density $N_{\rm{H}}$ by assuming that emissivity spectral index $\beta$ is 1.8, and dust opacity $\kappa$ is 0.1\,cm$^2$/g\,($\nu$/1000\,GHz)$^{\beta}$, which is assumed to be valid in high density environments~\citep{Hildebrand1983,Beckwith1990}.

\subsection{Planck} \label{sec:planck}

We use the polarized dust signal measured by the $Planck$-HFI 353\,GHz channel, where the signal-to-noise ratio (S/N) of the dust emission is maximum, as a tracer of the magnetic field (B). The $Planck$ observations provide Stokes $I$, $Q$, and $U$ parameter maps. The total polarized intensity $P$, polarization fraction $p$, polarization angle $\psi$, and plane of the sky (POS) magnetic field orientation angle $\theta_{\rm{B}}$ can be derived from the Stokes parameters using the equations

\begin{equation}
P = \sqrt{Q^2 + U^2},
\label{eq:P}
\end{equation}

\begin{equation}
p = P/I,
\label{eq:p}
\end{equation}

\begin{equation}
\psi = 0.5 \arctan(-U,Q),
\label{eq:psi}
\end{equation}

\begin{equation}
\theta_{\rm{B}} = \psi+\pi/2.
\label{eq:thetaB}
\end{equation}

In Eq.~\ref{eq:psi}, ${\rm{arctan}}(-U,Q)$ gives the angle ${\rm{arctan}}(-U/Q)$ in the correct quadrant, and the minus sign converts the $Planck$ data from {\tt{Healpix}} to IAU convention, where the polarization angle is counted positively from the Galactic North to the East. See, e.g., \citet{Planck2015XIX} for more details. 

Since our object, L1642, is located at high latitude, $Planck$ data have to be processed further to increase the S/N of the polarization quantities and to avoid any bias issues, as recommended by \citet{Montier2015I}. Hence the $Planck$ maps at 353\,GHz have been smoothed to 10{\arcmin} in order to reach a S/N $> 4$ in a region of a 30{\arcmin} radius centred on the cloud. Moreover, the modified asymptotic estimator \citep[hereafter MAS, ][]{Plaszczynski2014} has been chosen to provide a robust estimate of the polarization fraction, $p_{\rm{MAS}}$, as described in \citet[][]{Montier2015II}.
With this setup the uncertainty of the polarization angle remains below $5^{\circ}$ in the cloud and in most of the surrounding areas.

For the analysis of the regularity of the B field, we will also use
the polarization angle dispersion function, as defined in~\citet{Planck2015XIX}
with the following quantity:

\begin{equation}
S(\vect{x},\lag)= \left(\frac{1}{N} \sum_{i=1}^{N}{ (\DeltaAng_{xi})^2} \right)^{1/2}
\label{equ:delta_psi1}
\end{equation}
where $\DeltaAng_{xi}=\polang(\vect{x})-\polang(\vect{x}+\vect{\lag}_i)$ is the angle difference
between $\polang(\vect{x})$, the polarization angle at a given sky
position $\vect{x}$ (the central pixel), and $\polang(\vect{x}+\vect{\lag}_i)$ the
polarization angle at a sky position displaced from the centre
by $\vect{\lag}_i$.
The average is computed over an annulus of radius $\lag=|\vect{\lag}|$ and width
$\Delta\lag$. We use value 3.4$\arcmin$ for both radius and width.

\subsection{CO rotational emission}

We examine the dynamics of the cloud with the CO observations of~\citet{Russeil2003}, used also in~\citet{Malinen2014}. The SEST radio telescope mapped L1642 in the $J = 1-0$ and $J = 2-1$ transitions of $^{12}$CO, $^{13}$CO and C$^{18}$O. The half-power beam widths of the data are 45$\arcsec$ for $J = 1-0$ and 23$\arcsec$ for $J = 2-1$. The grid has 3$\arcmin$ spacing. Typical noise levels are $\Delta T_{\rm{rms}} = 0.06$ K for C$^{18}$O(2--1) and $\sim$0.15 K for other transitions.

\section[]{Methods} \label{sect:methods}

\subsection{Line Integral Convolution}  \label{sect:methods_LIC}

In order to visualize the B field orientation on other intensity maps, such as $Planck$ or {\it Herschel} emission maps, we use the Line Integral Convolution \citep[LIC,][]{Cabral1993} filtering technique. The method filters ('blurs') an input image texture along local vector field lines, in our case showing the orientation of the magnetic field lines.

When applying the LIC algorithm based on $Planck$ information at 10{\arcmin} on the 250\,$\mu$m {\it Herschel} map at $\sim$20{\arcsec} resolution, we make the assumption that the orientation of the B field is not changing inside the $Planck$ beam. This is a strong assumption which ignores any kind of small scale patterns, which could come for example from turbulence. Hence, this process is valid only when comparing large scale features.

\subsection{Rolling Hough Transform}   \label{sect:methods_RHT}

Several different methods have been used for identifying linear structures and quantifying their orientation. For example, \citet{Peretto2012} and \citet{Palmeirim2013} used the DisPerSE method~\citep{Sousbie2011} to extract the crests of linear structures, and calculated their average orientations. Other methods include column density gradients~\citep{Soler2013, Planck2015XXXV}, inertia matrix~\citep{Hennebelle2013}, and Hessian matrix~\citep{Planck2014XXXII}. 

As our interest here is mostly in diffuse striations, we use the Rolling Hough Transform~\citep[RHT,][]{Clark2014}. The RHT is a machine vision method based on the Hough transform~\citep{Hough1962}. It calculates the intensity as a function of angle in a circular region around every pixel of image data. The RHT thus quantifies the orientation of linear structures, rather than simply identifying them. By contrast, the DisPerSE method~\citep{Sousbie2011}, originally developed for cosmic web data, defines filaments as structures that connect local density maxima. This requirement is not well suited for characterizing striations. Indeed, we find that in order for DisPerSE to locate the faintest striations, the persistence threshold must be set so low that many spurious structures were also returned. Even then, the RHT-defined structures are a better representation of the striations seen by eye. 

We quantify the orientations of linear structures, and the orientation uncertainties, from the RHT output using the equations defined in~\citet{Clark2015}. The RHT approach enables a direct pixel-by-pixel comparison of the structure orientation with the magnetic field orientation. We use the default parameter values of the RHT.

The RHT output is three-dimensional: intensity as a function of angle for every pixel in the input image. Integrating the RHT output over all possible angles produces the RHT backprojection, a visualization of the linear intensity in the image \citep[see][for details]{Clark2014}.

\subsection{Histogram of Relative Orientations}  \label{sect:methods_HRO}

\citet{Soler2013} \citep[and][]{Planck2015XXXV} use the name Histogram of Relative Orientations (HRO) to refer to their method of using column density gradients to extract linear structures. Confusingly, they use the same name to refer to the end result, the actual histogram comparing the relative orientation of structures and magnetic field. We take the same approach as~\citet{Planck2015XXXVIII} and use the name HRO to refer just to the histogram in general, regardless of how the orientation of the structures has been determined.

We perform an HRO analysis to quantify the alignment between the magnetic field and the {\it Herschel} 250\,$\mu$m elongated structures traced by the RHT method. We build the angle difference between the B field orientation $\theta_{\rm{B}}$ and the structure orientation $\theta_{\rm{RHT}}$ in each pixel by
\begin{equation}
\Delta \theta = \theta_{\rm{B}} - \theta_{\rm{RHT}} \, .
\end{equation}
This angle difference is only computed where the uncertainty of the B field and the RHT orientations are simultaneously 
below $10^{\circ}$.

\subsection{Multivariate analysis methods for analyzing HROs}  \label{sect:methods_NMF}

We want to examine how the relative alignment of dust structure and magnetic field changes with column density. It is expected that the two are preferentially aligned in low column density environments, and perpendicular in high column densities~\citep[e.g.,][]{Planck2015XXXV}.
We want to go a step further to build a robust quantification to reveal the transition between the two modes, without any a priori knowledge on the components. This is why we use multivariate analysis methods to determine the basic ways in which the HROs vary as a function of column density.

We use a two step method to analyze quantitavely the HROs described in the previous section. In the first step, we use a variant of 
Principal Component Analysis~\citep[PCA, see e.g.,][]{Jolliffe2002} to estimate the number of components in the HROs. 
PCA is a statistical method used to find the principal components of the data, i.e., the directions with the most variance. The method is often used to reduce the dimensions of the data. First, we divide the $Herschel$ map into 18 continous bins of column density, all containing the same number of pixels. The HRO analysis described above is then performed on each column density bin separately, yielding a set of 18 HROs. Then, we create the matrix $M$, whose columns contain the 18 HROs. We compute the principal components of matrix $M$ using PCA. The principal components constitute a basis to the set of histograms contained in $M$. Since the components are orthogonal by definition it is difficult to interpret them in physical terms (in particular because they contain negative elements). Therefore, we only use PCA to estimate the number of main components.

In the second step we apply non-negative matrix factorization (NMF, \citealt{lee99}) to the matrix $M$. NMF is a multivariate analysis technique, which is used to factorize a matrix $M$ of dimension ($m \times n$) into two matrices $W$ ($m \times p$) and $H$ ($p \times n$), where $p$ is usually notably lower than $m$ or $n$. All the matrices can have only non-negative elements. Usually, the problem must be approximated numerically forming $V = W \times H + U$, where $U$ is a residual matrix.
The dimensions of $W$ and $H$ must be defined. $m$ is defined by the number of rows in $M$, i.e., the number of different angle bins in our case. $n$ is defined by the number of columns in $M$, i.e., the number of HROs ($n=18$) in our case, while $p$, the number of rows in $H$, or elementary histograms we wish to extract, must be set by the user. We use the number of the main components in $M$ obtained with PCA in step one. The algorithm to compute $W$ and $H$ was presented in \citet{lee99}. It is initialized with random $W$ and $H$ matrices and it uses iterative update rules to minimise the Euclidian distance between $M$ and $W\times H$. It should be noted that NMF is similar in many regards to ``positive matrix factorization" (PMF), introduced by \citet{paa94}, which differs mainly in terms of optimization algorithm. Both NMF and PMF have been applied in astronomy~\citep[see, e.g.,][]{ber07,juv96}.

The solution to NMF is not unique and can depend on the initialization, hence we run NMF 1000 times on $M$ to empirically verify that the solution is constant in our specific case. The resulting components in $H$ extracted by NMF are elementary histograms. The NMF method produces directly the weights contained in $W$, i.e., the contribution of each component to the histograms of $M$, which can then be shown as a function of the column density. This analysis is performed in Sect.~\ref{sect:hro}.

\section[]{Results} \label{sect:results}

\subsection{$Planck$ polarization and magnetic field orientation}

\begin{figure*}
\includegraphics[width=12cm]{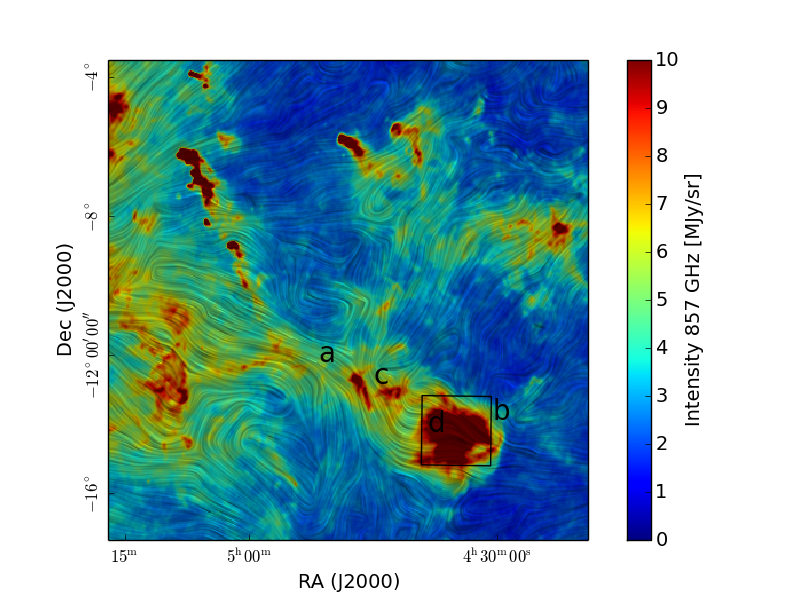}
\caption{Region surrounding L1642 in $Planck$ 857\,GHz intensity map with POS magnetic field orientation shown by the LIC texture at 30$\arcmin$ resolution (see main text for details). Structures discussed in the text are marked with letters $a-d$. L1642 ($\alpha_{2000} = 4^{\rm h}35^{\rm m}$ and $\delta_{2000} = -14^{\circ}15'$) is located at the SW corner of the image (marked with a square), at the head of an over 5$^{\circ}$ long HI pillar oriented towards NE. There are also two smaller dense blobs at the center of the pillar, at approximately $\alpha_{2000} = 4^{\rm h}45^{\rm m}$, marked with $c$. The details of the figure are best seen in the electronic version.}
\label{fig:planck_large}
\end{figure*}

The $Planck$ 857\,GHz intensity map with POS magnetic field orientation of the area surrounding L1642 is shown in Fig.~\ref{fig:planck_large}. 
L1642 is located at the head of a large, over 5$^{\circ}$ long pillar structure visible in HI and infrared/submillimeter dust emission maps. We use the LIC method described in Sect~\ref{sect:methods_LIC} to visualize the POS magnetic field. The magnetic field orientation is shown at 30$\arcmin$ resolution, since, at full resolution, the S/N of the polarization data in the surrounding low column density region is low, as there is not much material to emit polarized radiation. We use Equatorial coordinates and refer to directions North (N, up), South (S, down), East (E, left), and West (W, right) in this and all other images.

The large scale POS magnetic field in Fig.~\ref{fig:planck_large} shows a lot of turmoil in this region. The areas discussed below are marked with letters $a-d$ in the figure. The large pillar leading to L1642 appears to be a turning point to the large scale magnetic fields. On the E part of the pillar ($a$), the magnetic field is oriented approximately midway between the E-W and NE-SW directions. On the NW side of L1642 ($b$) the B field is oriented in the NW-SE direction, that is approximately perpendicular to the E side. In the center of the pillar ($c$), where there are two dense, elongated structures, these perpendicular B fields meet and the field lines bend. However, at the W part of the pillar and at the head where L1642 is situated ($d$), the B field is again at E-W orientation. Thus, the NW part of L1642 faces again a perpendicular B field. The impact of the surrounding region to the L1642 cloud is discussed more in Sect.~\ref{sect:discussion}.

\begin{figure*}
\begin{tabular}{ccc}
\includegraphics[width=0.43\textwidth]{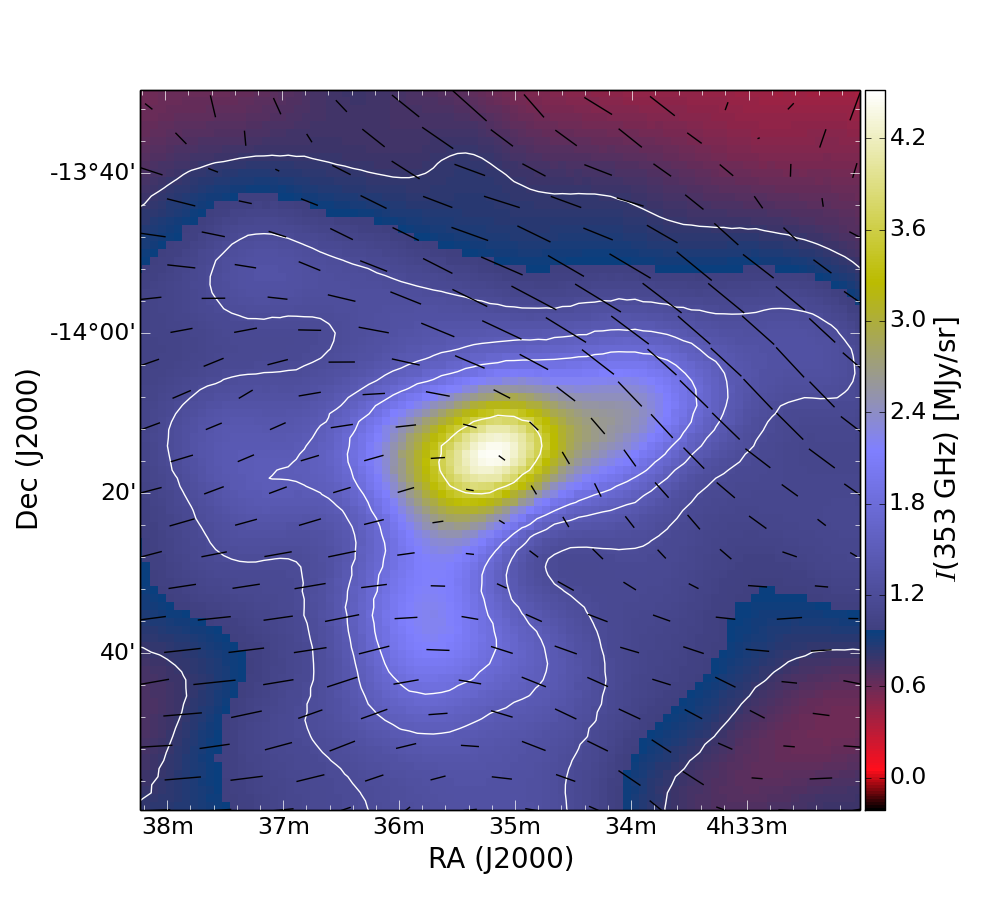}
\hspace{30pt}
\includegraphics[width=0.5\textwidth]{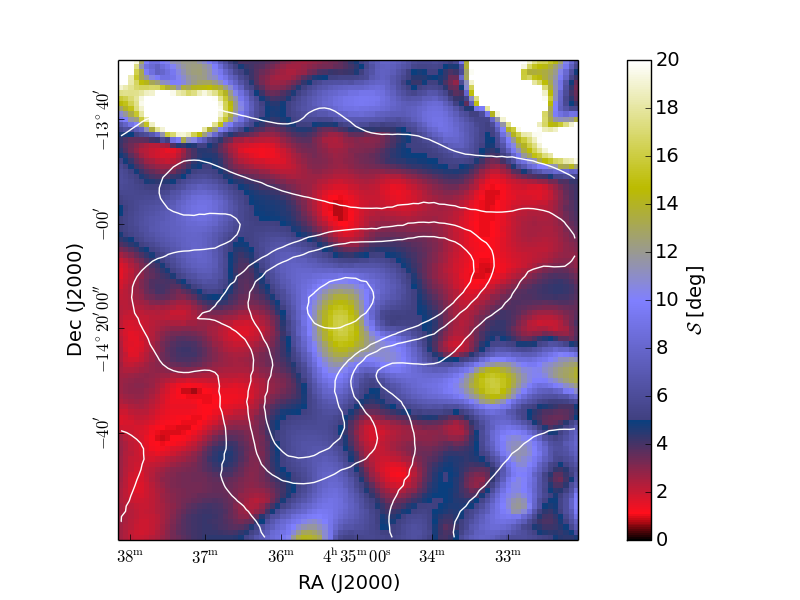}\\
\includegraphics[width=0.5\textwidth]{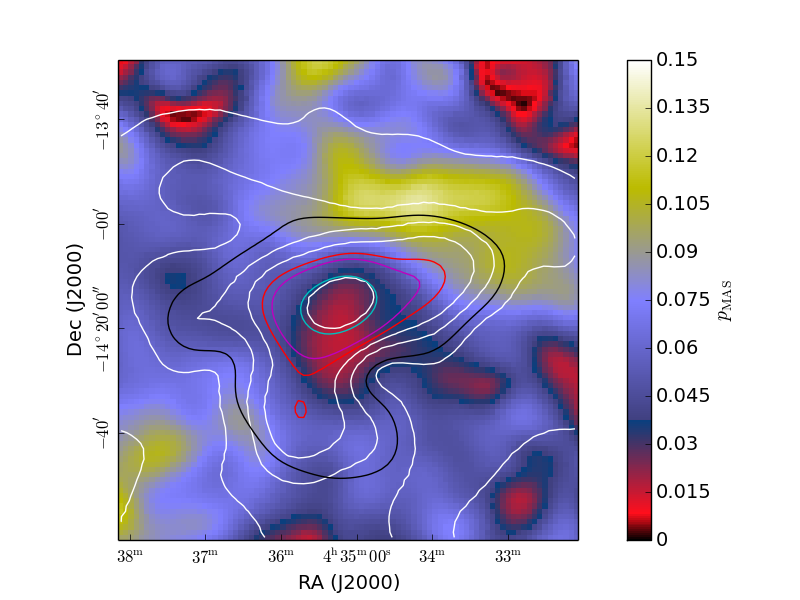}
\includegraphics[width=0.5\textwidth]{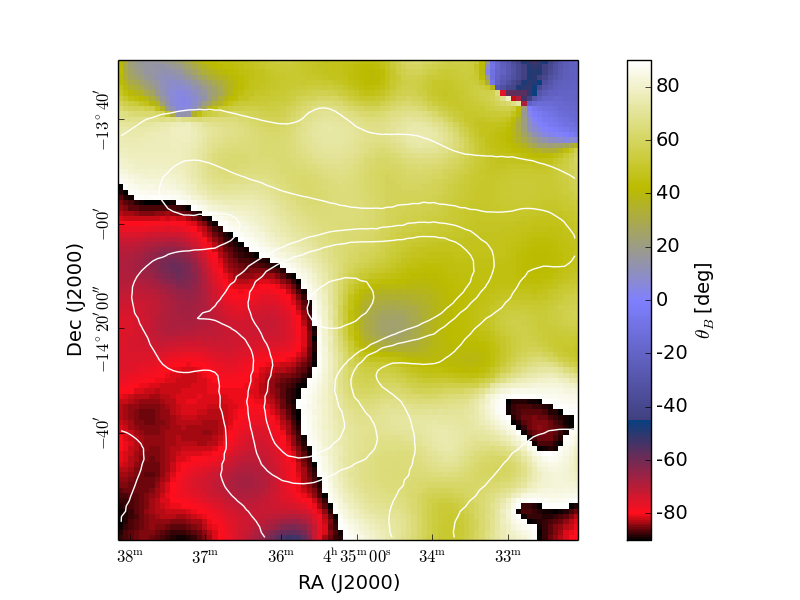} \\
\includegraphics[width=0.5\textwidth]{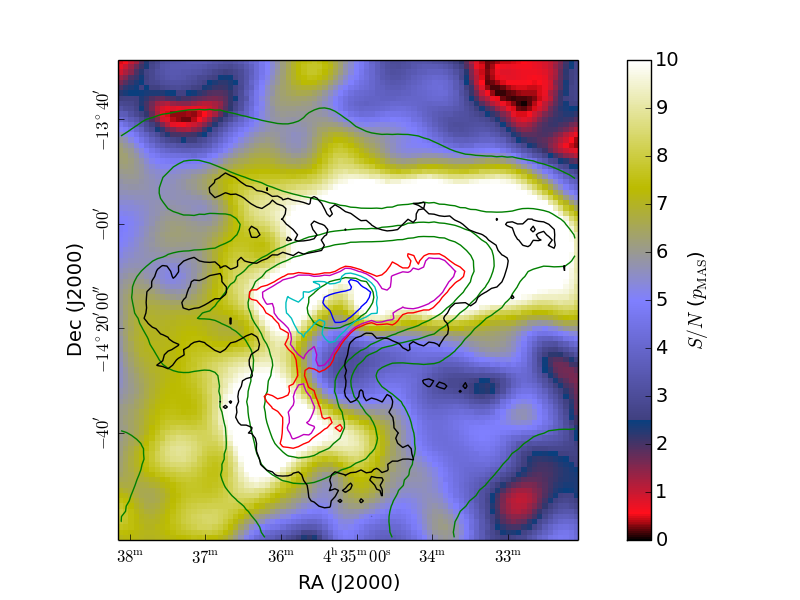}
\includegraphics[width=0.5\textwidth]{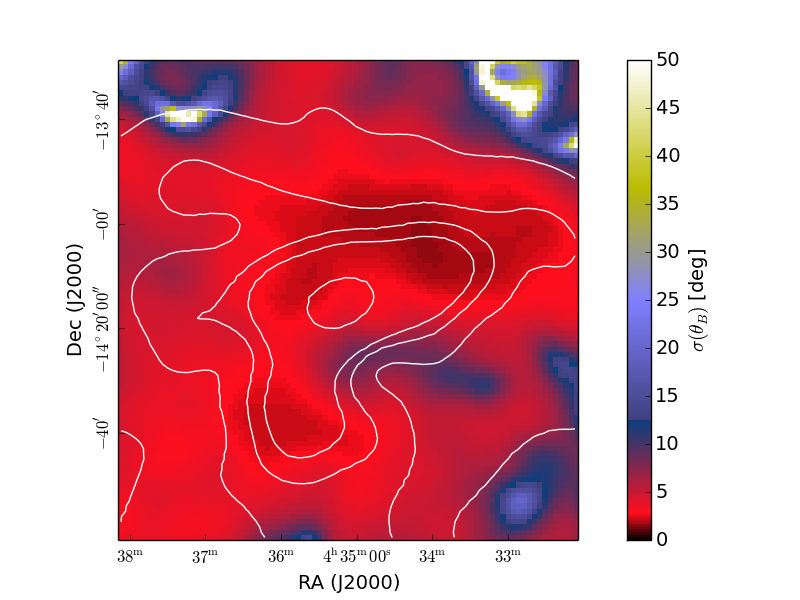} \\
\end{tabular}
\caption{$Planck$ 353\,GHz intensity, polarization angle dispersion function $S$, MAS estimate of polarization fraction $p_{\rm{MAS}}$, magnetic field orientation angle $\theta_{\rm{B}}$, S/N of $p_{\rm{MAS}}$, and $\sigma(\theta_{\rm{B}})$ maps at 10$\arcmin$ resolution. The white (and green) contours are from the $Planck$ 353\,GHz intensity map. B field is visualized with vectors on the intensity map. The blue, cyan, magenta, red, and black contours on the $p_{\rm{MAS}}$ and S/N of $p_{\rm{MAS}}$ maps show the column density based on $Herschel$ data at 5.22, 3.0, 2.0, 1.74, and 1.0 $\times 10^{21}\, {\rm{cm}}^{-2}$, respectively. The $Herschel$ data are shown at 10$\arcmin$ resolution (on the $p_{\rm{MAS}}$ map) and at the original 40$\arcsec$ resolution (on the S/N of $p_{\rm{MAS}}$ map). The highest (blue) contour is not seen at 10$\arcmin$ resolution.}
\label{fig:planck}
\end{figure*}

Fig.~\ref{fig:planck} shows maps of the $Planck$ 353\,GHz intensity $I$, polarization angle dispersion function $S$, the MAS estimate of polarization fraction $p_{\rm{MAS}}$, magnetic field orientation angle $\theta_{\rm{B}}$, S/N of the $p_{\rm{MAS}}$, and the dispersion of the orientation angle $\sigma(\theta_{\rm{B}})$ of the L1642 cloud, all at 10$'$ resolution. The magnetic field orientation $\theta_{\rm{B}}$ shows a sharp transition between the NW and SE regions, as the B field orientation changes abruptly near the densest part of L1642.

The polarization fraction $p_{\rm{MAS}}$ is very low at the densest parts of the cloud. The polarization fraction is high only in the N part of L1642, where striations are more visible. There is clear anti-correlation between $S$ and $p_{\rm{MAS}}$. In the N area of the striations, the $S$ level is low, and $p_{\rm{MAS}}$ relatively high, which is consistent with a well ordered magnetic field in this part.
As shown in the Planck observations overview~\citep{Planck2015XIX} and related simulations~\citep{Planck2015XX}, variations of the B field direction within the beam or along the line of sight (e.g. due to tangling), would result in a decrease of the observed polarization fraction. The relatively high value of $p_{\rm{MAS}}$, coinciding with a low $S$ level indicates that the field is uniform in this area.

\subsection{Structures in {\it Herschel} submm maps}
\label{sect:rht}

\begin{figure*}
\includegraphics[width=0.49\textwidth]{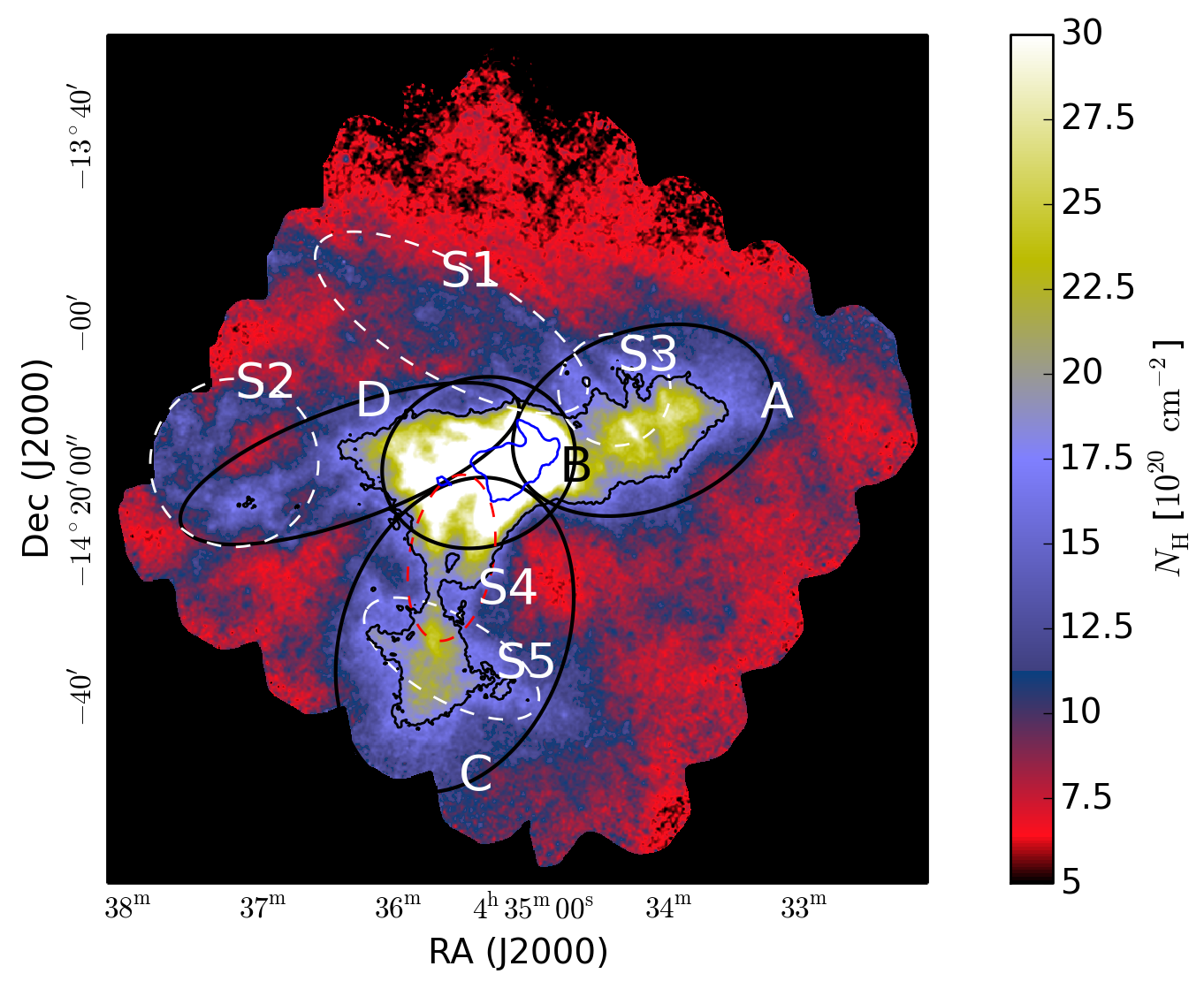}
\includegraphics[width=0.49\textwidth]{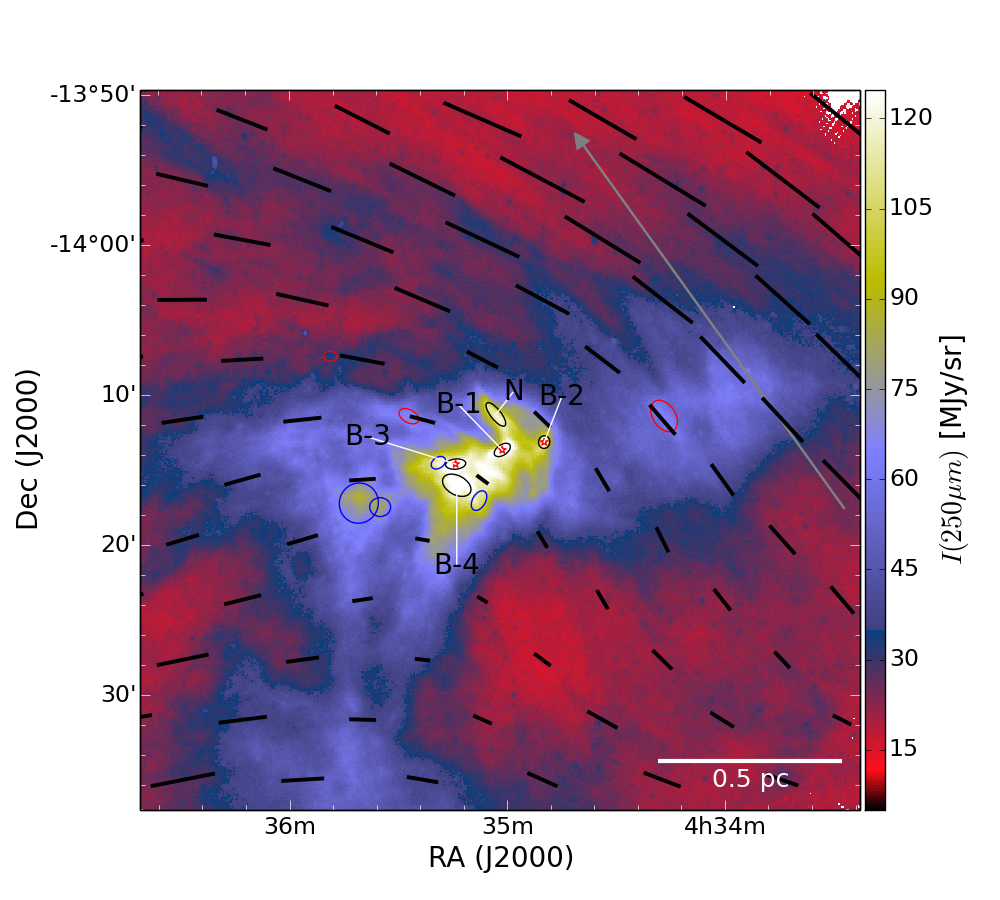}
\caption{($Left$) {\it Herschel} column density map with regions A, B, C, and D, and contours at $1.74 \times 10^{21}\, {\rm{cm}}^{-2}$ and $5.22 \times 10^{21}\, {\rm{cm}}^{-2}$ used for RHT analysis (see text for details). Other structures discussed in the text are marked with dashed white (or red) ellipses. S1, S2, and S3 mark striations, S4 marks elongated structures between regions B and C, and S5 marks another linear feature almost perpendicular to the S4 structures. ($Right$) {\it Herschel} 250 $\mu$m intensity map with sources from~\citet{Malinen2014}: YSOs B-1, B-2, and B-3, cold clump B-4, and an elongated structure N marked with black ellipses. The YSOs are also marked with red stars. Bound and unbound submillimetre clumps from~\citet{Montillaud2015} are marked with blue and red ellipses, respectively. B field is visualized with vectors. The grey arrow shows the location of the cut used in the CO analysis in Fig.~\ref{fig:CO}.}
\label{fig:herschel}
\end{figure*}

\citet{Malinen2014} examined the general structure and star formation activity of molecular cloud L1642 using especially $Herschel$ dust emission maps. Their data showed differences in the different parts of the cloud. In this paper, we continue with a more detailed analysis of the cloud structure. Fig.~\ref{fig:herschel} ($left$) shows the $Herschel$ column density map with named regions A, B, C, and D. Although the cloud is shaped more like a blob than a filamentary structure, it contains several elongated features. As noted in~\citet{Malinen2014}, there are clear striations in the diffuse cloud surrounding the star-forming, dense area, e.g, in the N side of the cloud (S1 in the figure). There are also interesting structures that may fall between the concepts of "filament" and "striation", for example the spiraling "tail" in the East (S2 in the figure). There are also striations directly connected to the main parts of the cloud, especially in the W part, where they form part of the A structure (S3). All these striations have column densities of typically  $N_{\rm{H}} < 2 \times 10^{21}\, {\rm{cm}}^{-2}$. Also, the two elongated, straight structures (S4) between regions B and C are not typical filaments or striations. S5 marks another linear feature almost perpendicular to the S4 structures.

The previously known objects, including three Young Stellar Objects (YSOs) binary B-1, binary B-2, and B-3, and a cold clump B-4, inside the densest cloud \citep{Malinen2014}, together with the submillimetre clumps from~\citet{Montillaud2015} have been marked in Fig.~\ref{fig:herschel} ($right$). Galaxies and non-reliable objects have been removed from the list. In the densest central part of the cloud, there is also an elongated extension, resembling a finger (marked N in the figure), towards North. This type of structure would not usually be considered filament or striation.

To extract structures in the $Herschel$ map, we use the RHT method (described in Sect.~\ref{sect:methods_RHT}). We analysed the RHT results separately for the whole area and for selected regions. We used the column density $N_{\rm H}$ map to define a threshold in column density to separate diffuse
($N_{\rm{H}} < 1.74 \times 10^{21}\, {\rm{cm}}^{-2}$)
and dense ($N_{\rm{H}} > 1.74 \times 10^{21}\, {\rm{cm}}^{-2}$)
regions. We then divided the dense medium into four regions: A, B, C, and D. The region B is defined as the region of highest column density, 
$N_{\rm{H}} > 5.22 \times 10^{21}\, {\rm{cm}}^{-2}$, 
shown in Fig.~\ref{fig:herschel} ($left$) with a contour. The other regions are defined as belonging to the respective ellipse shown in Fig.~\ref{fig:herschel} ($left$), and having column density values $1.74 \times 10^{21}\, {\rm{cm}}^{-2} < N_{\rm{H}} < 5.22 \times 10^{21}\, {\rm{cm}}^{-2}$.

\begin{figure}
\begin{tabular}{c}
\includegraphics[width=0.48\textwidth]{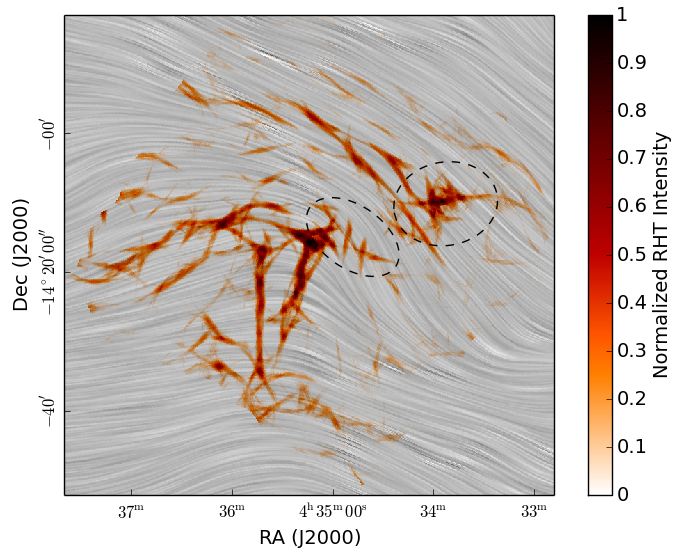}  \\
\includegraphics[width=0.48\textwidth]{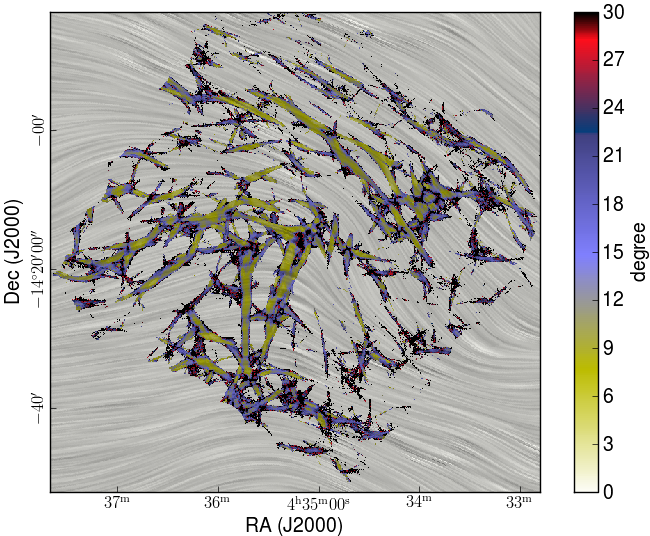}  
\end{tabular}
\caption{($Top$) Structures extracted by RHT analysis of {\it Herschel} 250\,$\mu$m map. The ellipses show two structures resembling a fishbone with "spine" and approximately perpendicular "bones". ($Bottom$) Uncertainty of the RHT orientation angle $\theta_{\rm{RHT}}$. The HRO analysis is restricted to regions where the uncertainty is below 10$^{\circ}$. The grey background shows the POS magnetic field orientation with LIC texture.}
\label{fig:rht}
\end{figure}

Fig.~\ref{fig:rht} shows the linear stuctures extracted with the RHT method, and the reliability of the detections. The results match the structures seen in the map very well, revealing detailed, linear structures both in the diffuse regions and in the denser parts of the cloud. We note that in addition to these, the RHT map shows two structures resembling the body of a fishbone, with linear "spine" and approximately perpendicular "bones", marked in the Fig.~\ref{fig:rht} ($top$). The reliability map does not show notable difference in the structures between diffuse and dense regions.

\subsection{Visual comparison of $Planck$ and $Herschel$ data}

\begin{figure*}
\includegraphics[width=12cm]{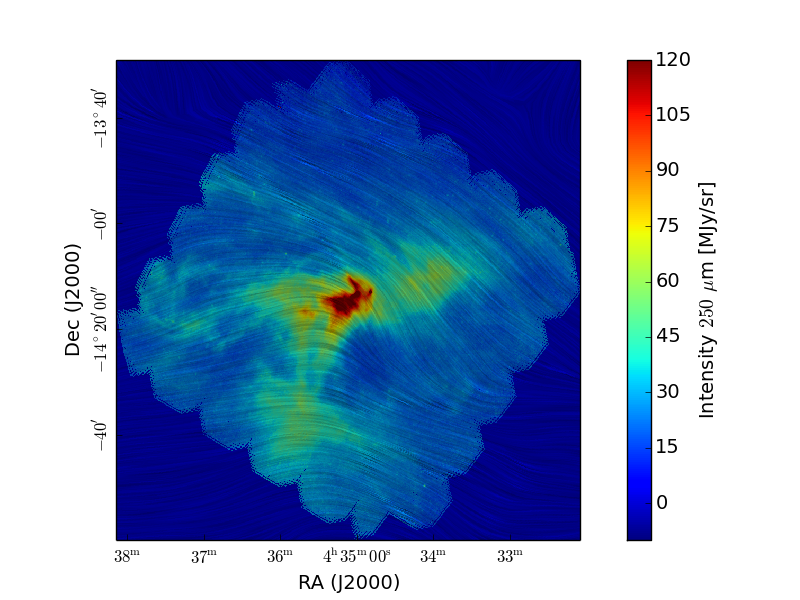}
\caption{L1642 in $Herschel$ 250 $\mu$m map (at 18.3$''$ resolution) with $Planck$ magnetic field orientation (at 10$'$ resolution) visualized by LIC texture. The details of the figure are best seen in the electronic version.}
\label{fig:herschelplanck}
\end{figure*}

Fig.~\ref{fig:herschel} ($right$) shows the magnetic field lines derived from $Planck$ polarization map (at 10$'$ resolution) plotted with vectors on the $Herschel$ 250 $\mu$m intensity map (at $\sim$18.3$''$ resolution) of the cloud L1642. The length of the vectors are proportional to the polarization fraction. For comparison, Fig.~\ref{fig:herschelplanck} shows the B fields with a LIC map (described in Sect.~\ref{sect:methods_LIC}). There are clear differences in the magnetic field orientations in the different areas of the cloud. In the NW part of the cloud, the magnetic field lines are oriented between NE-SW. 
In the S and E parts of the cloud, the lines are oriented mostly between E-W. However, when the lines reach the W point of the dense region, they turn abruptly approximately 90$^{\circ}$, almost towards S, along with the lines in the W part. Similarly, when the lines at the E and N regions meet at NE at the outskirts of the cloud, they turn towards the same orientation, almost to S. The densest part of the cloud is exactly where the B field is bending the most.

The Eastern, filamentary curving tail (S2) of the cloud is located in the area where B field lines with different orientations meet. 
The orientation of the two linear structures (S4) between the regions B and C differs by $\sim$30$^{\circ}$.
As the field lines curve in this region, both structures are in fact approximately perpendicular to the local magnetic field.
The largest elongated structure (N) of the main cloud pointing towards NE is approximately aligned with the magnetic field.
Some of the structures extracted by the RHT method, especially in region A and West of region B, resemble fishbones, with an elongated structure as the spine, and perpendicular smaller structures or striations on both sides. Comparing to the LIC texture, these structures are not exactly perpendicular or aligned with the magnetic field, but could be moving and evolving. Especially region A looks as if it is bend and shaped by the magnetic field. The roundish general structure is revealed to be formed of thin slices when looked in more detail.
Comparing the magnetic field lines to the $Herschel$ map, it is evident how well the field lines follow the general structure of the cloud even to the finer details, and the diffuse striations surrounding the cloud.

\subsection{Polarization fraction versus column density}

\begin{figure}
\includegraphics[width=8.5cm]{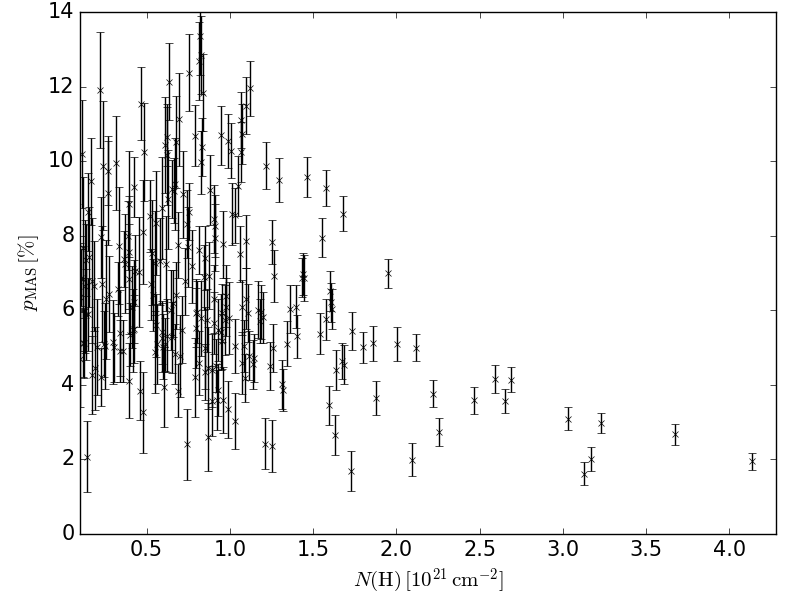}
\caption{Polarization degree $p_{\rm{MAS}}$ (based on $Planck$ data) as a function of the column density (based on $Herschel$ data). The data are at a resolution of 10$\arcmin$, and the samples are taken every 5$\arcmin$.}
\label{fig:pvsNH}
\end{figure}

Fig.~\ref{fig:pvsNH} shows the polarization fraction $p_{\rm{MAS}}$ as a function of the column density (derived from $Herschel$ maps), for data at 10$\arcmin$ resolution. The sample points are taken every 5$\arcmin$. The polarization fraction is notably reduced above $N_{\rm{H}} = 2 \times 10^{21}\, {\rm{cm}}^{-2}$. The upper envelope of the polarisation fraction distribution begins to decrease already below this column density value. Reduced polarization fraction is observed over a large area where the magnetic field orientation is still mainly uniform (see Fig.~\ref{fig:planck}). For comparison, the middle left frame of Fig.~\ref{fig:planck} shows the column density contours (at 10$\arcmin$ resolution) on the $p_{\rm{MAS}}$ map. Although most of the high column density pixels originate in the central region B, also region A has a significant number of pixels above $N_{\rm{H}} = 2 \times 10^{21}\, {\rm{cm}}^{-2}$. All of the highest column density pixels $N_{\rm{H}} > 3 \times 10^{21}\, {\rm{cm}}^{-2}$ originate in the central region B within a few $Planck$ resolution elements. The high column density points in Fig.~\ref{fig:pvsNH} are therefore highly correlated and come from a relatively small area.

Part of the depolarisation (and most of the scatter) could be related to changes in magnetic field orientation, either at small scales not resolved by $Planck$ or in regions along the line of sight where POS orientation of the magnetic field differs. However, it is possible that the column density dependence is caused by a partial loss of grain alignment in the densest parts of the cloud. If grain alignment is assumed to be caused by radiative torques~\citep[see review by][]{Lazarian2007}, the change would be directly related to the attenuation of the radiation field. This increases the lower size limit of the dust grains that remain aligned in the magnetic field. Thus, Fig.~\ref{fig:pvsNH} could point to an almost complete loss of grain alignment in the cloud center and the minimum values of $p_{\rm MAS}$ could be attributed to polarised emission from diffuse regions along the line of sight.

The column density dependence of the polarization fraction has been discussed in the recent $Planck$ papers~\citep{Planck2015XIX, Planck2015XX, Planck2014XXXIII}. \citet{Planck2015XIX} concluded that the general decrease of $p$ with $N_{\rm{H}}$, at the resolution of 1$^{\circ}$, was mainly due to fluctuations in the B field orientation along the line of sight, probing various components in particular toward regions close to the galactic plane. This was supported by the anti-correlation observed between $p$ and $S$. In our case of a high-latitude cloud, there is not as much confusion along the LOS. However, we see an increase of $S$ toward the central part associated with the highest $N_{\rm{H}}$ values. In this paper, we are only looking at one cloud and not making a statistical analysis of several regions. Therefore, it is not clear what is the cause of the decrease of $p$ in our particular case.

\subsection{Relative orientation of magnetic fields and $Herschel$ structures}
\label{sect:hro}

\subsubsection{Relative orientation in different regions}

\begin{figure}
\begin{tabular}{c}
\includegraphics[width=0.5\textwidth]{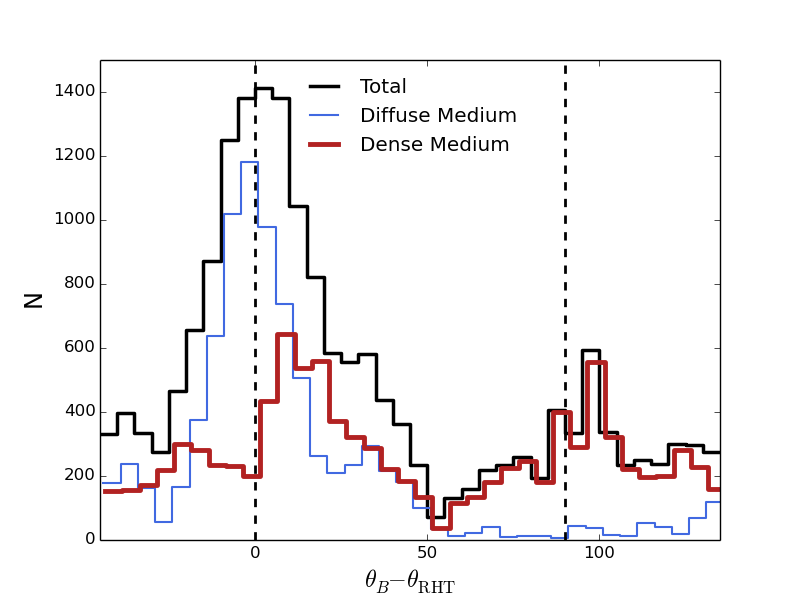}  \\
\includegraphics[width=0.5\textwidth]{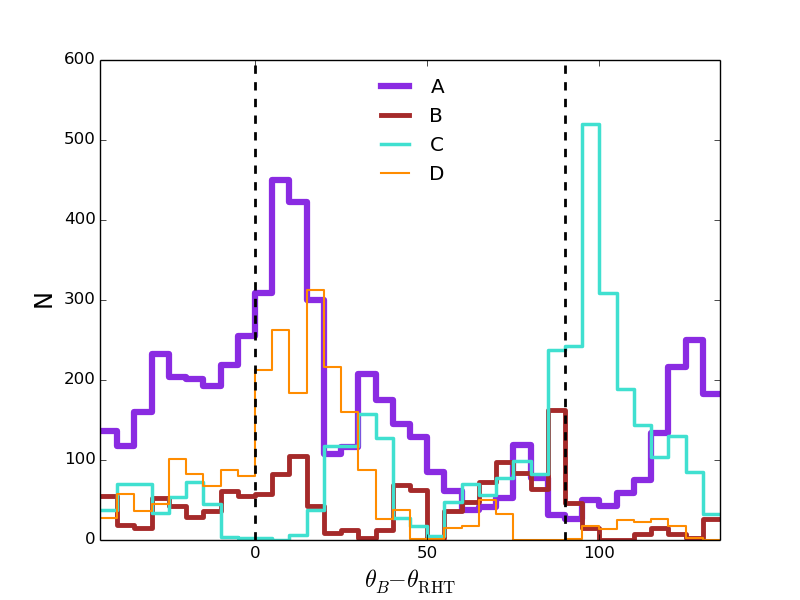}  \\
\end{tabular}
\caption{HRO analysis comparing the relative orientation of magnetic field and cloud structure in the case of the whole data or dense and diffuse areas ($top$), and in the case of separate regions defined in the main text and shown in Fig.~\ref{fig:herschel} ($left$). $N$ gives the number of pixels. Vertical dashed lines show the values 0$^{\circ}$ and 90$^{\circ}$. The lines for dense and diffuse areas in the ($top$) figure are shifted slightly to the right.}
\label{fig:hro}
\end{figure}

We perform a Histogram of Relative Orientations (HRO) analysis (as described in Sect.~\ref{sect:methods_HRO}) to quantify the alignment between the magnetic field and the {\it Herschel} 250\,$\mu$m elongated structures traced by the RHT method. The resulting histograms are shown in the top frame of Fig.~\ref{fig:hro} for the entire region, and split between the diffuse and dense regions. There are two peaks in this distribution, approximately centred at $0^{\circ}$ and $90^{\circ}$. The HRO computed on the diffuse regions is mainly associated to the first peak at $0^{\circ}$, while the HRO of the denser regions is more complex.

When splitting the denser areas into four physical regions, A, B, C, and D (as defined in Sect.~\ref{sect:rht}), it appears that the structures in C are clearly perpendicular to the magnetic field, while the A and D regions have structures mainly aligned with the magnetic field (see Fig.~\ref{fig:hro} lower frame). In the densest region B, there is no clear distinction anymore. This is mainly due to the fact that the distribution of the RHT angles (tracing the matter) is much more flat because of confusion along the line of sight. Fig.~\ref{fig:planck} shows that the orientation of the magnetic field is reliable and very uniform in the full region. We also note that the distribution peaks are not exactly at $0^{\circ}$ and $90^{\circ}$, but slightly above those values. Region A also shows another, smaller peak at $\sim130^{\circ}$. The "fishbone" like structures shown in Fig.~\ref{fig:rht} ($top$) are in the A region, and are likely to contain most of the pixels causing this peak.

\subsubsection{Column density dependence of relative orientation}  \label{sect:nh}

\begin{figure}
\includegraphics[width=0.5\textwidth]{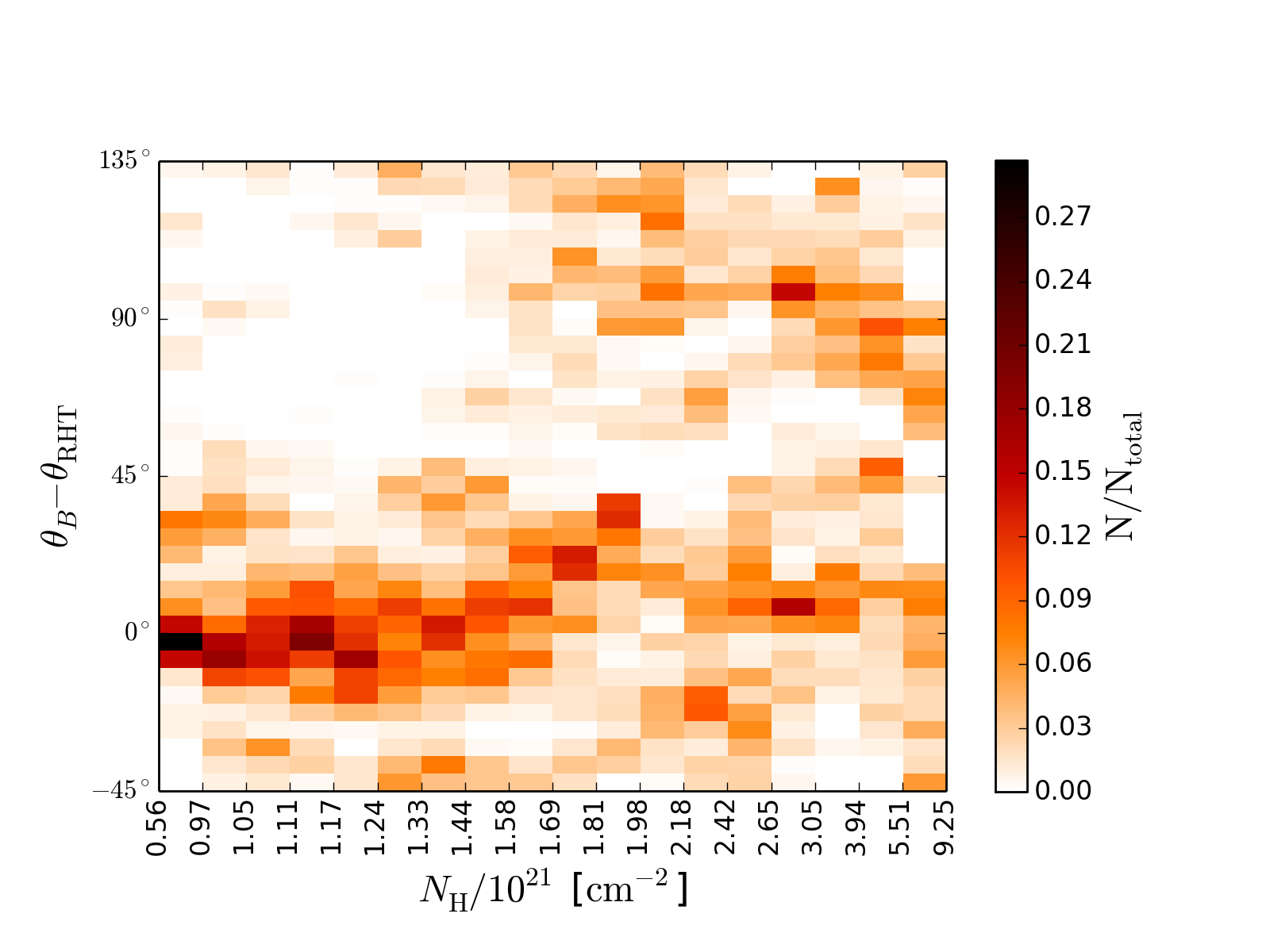}
\caption{The elementary histograms of angle difference built per bin of column density, so that all histograms contain the same number of pixels. Each histogram is normalised by its integral.}
\label{fig:hro_nh}
\end{figure}

We analyze the column density dependence of relative orientation using the method described in Sect.~\ref{sect:methods_NMF}.
We have divided the $Herschel$ map into 18 continuous bins of column density between $0.56 \times 10^{21}\, {\rm{cm}}^{-2}$ and $9.25 \times 10^{21}\, {\rm{cm}}^{-2}$, all containing the same number of pixels. The HRO analysis described above is then performed on each column density bin separately, yielding a set of 18 HROs, which are combined in matrix $M$, and shown in Fig.~\ref{fig:hro_nh}. We run an initial PCA analysis on this set, and find that 95\% of the power in $M$ can be expressed with the first two components.

\begin{figure}
\includegraphics[width=0.5\textwidth]{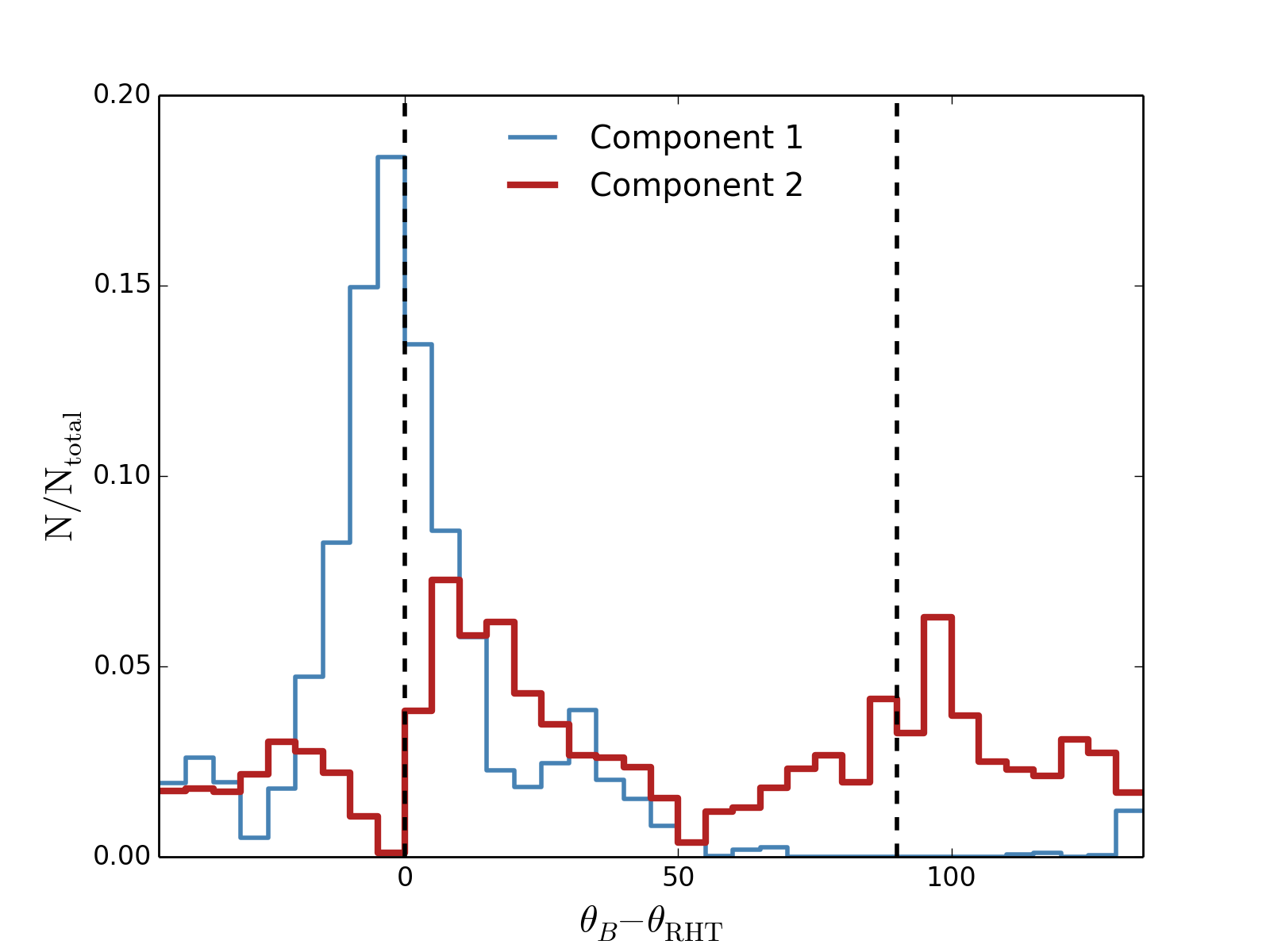}
\caption{The two principal components found with the NMF algorithm applied on the 18 histograms of angle difference built per column density bin. The horizontal axis shows the relative orientation of magnetic field and cloud structure. Each component is normalised by its integral.
}
\label{fig:pca_nh}
\end{figure}

In the second step, we apply NMF to factorize matrix $M$ into the product $W\times H$, where the dimensions of $H$ are ($p \times n$), and $n = 18$. As we have seen above in the PCA analysis, the data is contained in a dimension 2 subspace and hence we can apply NMF directly with $p=2$. We then run NMF 1000 times on $M$ to empirically verify that the solution is constant. The two final averaged components found in the Monte Carlo runs are shown in Fig.~\ref{fig:pca_nh}. As expected, one component has a clear peak at 0$^{\circ}$, while the other component has two main peaks, near 0$^{\circ}$ and 90$^{\circ}$.

\begin{figure}
\includegraphics[width=0.5\textwidth]{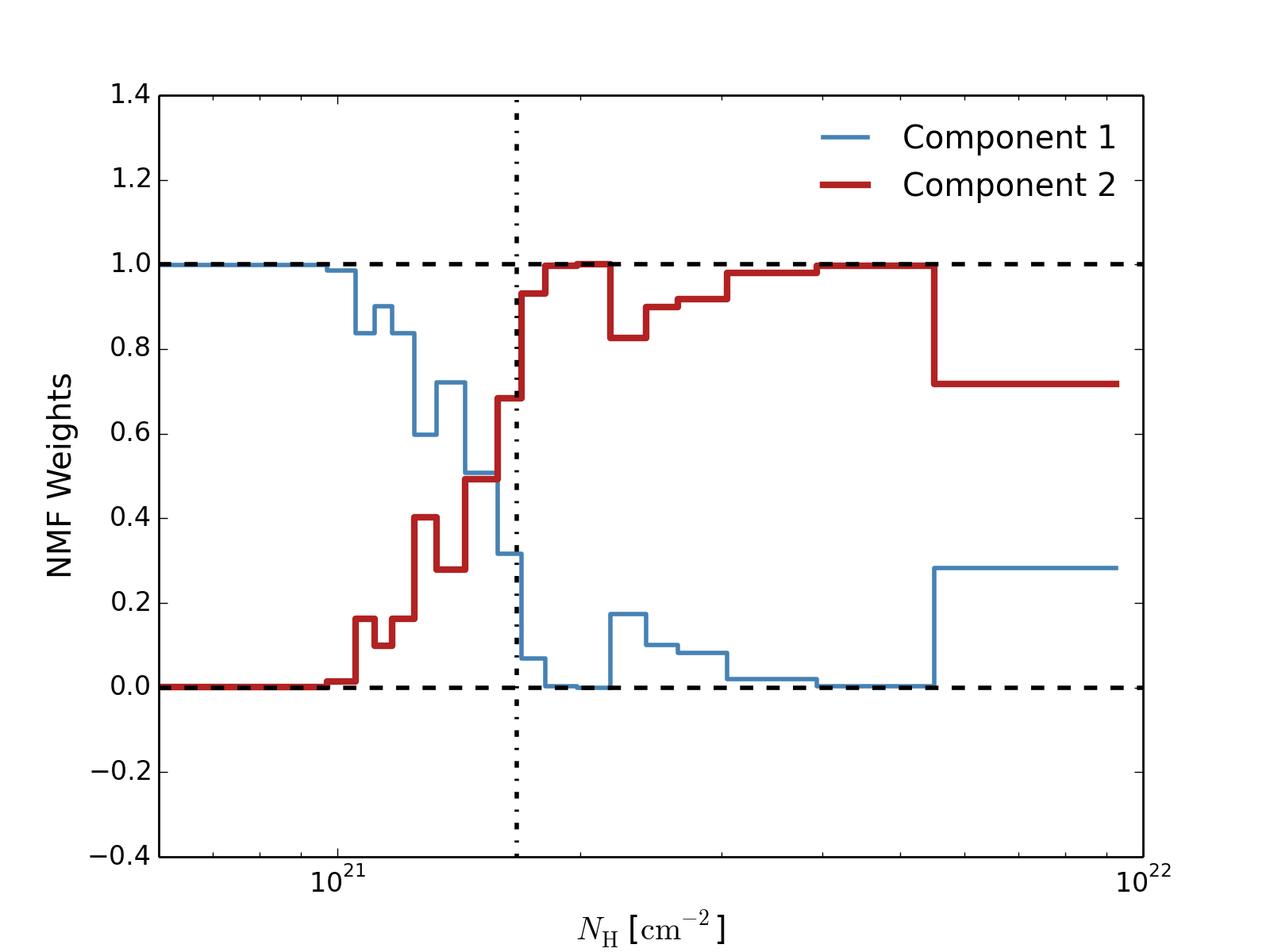}
\caption{The weights of the components in NMF analysis, i.e., the contribution of each component to the 18 histograms as a function of column density. The image shows the transition where the relative orientation of density structures and magnetic field lines changes from preferentially aligned to perpendicular. The dashed vertical line shows the similar result obtained by~\citet{Planck2015XXXV}, see main text for details.}
\label{fig:nmf_nh}
\end{figure}

The weights contained in $W$, i.e. the contribution of each component to the 18 histograms of $M$ are shown as a function of the column density in Fig.~\ref{fig:nmf_nh}. This confirms that one component is associated with diffuse regions and one with dense regions.
We see a clear transition around a column density at $N_{\rm{H}} \sim 1.6 \times 10^{21}\, {\rm{cm}}^{-2}$. \citet{Planck2015XXXV} derived a threshold value of $N_{\rm{H}} = 10^{21.7} {\rm{cm}}^{-2} \sim 5 \times 10^{21} {\rm{cm}}^{-2}$ in their analysis, but using another convention for the dust opacity $\kappa$. To compare with our estimate, we need to divide their value by 3 leading to $N_{\rm{H}} \sim 1.67 \times 10^{21}\, {\rm{cm}}^{-2}$, which is in good agreement with our result. The other way, our estimate corresponds to $N_{\rm{H}} \sim 4.8 \times 10^{21}\, {\rm{cm}}^{-2} \sim 10^{21.7} {\rm{cm}}^{-2}$ using their convention for $\kappa$.

\subsection{Comparison to CO} \label{sect:CO_results}

\begin{figure*}
\begin{tabular}{ccc}
\includegraphics[width=0.5\textwidth]{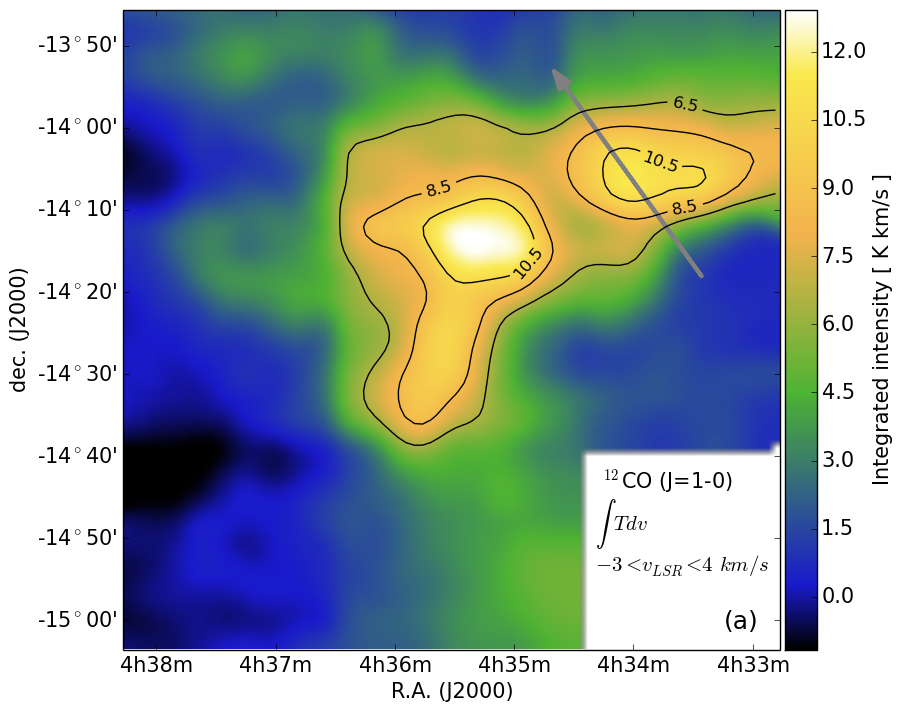}
\includegraphics[width=0.5\textwidth]{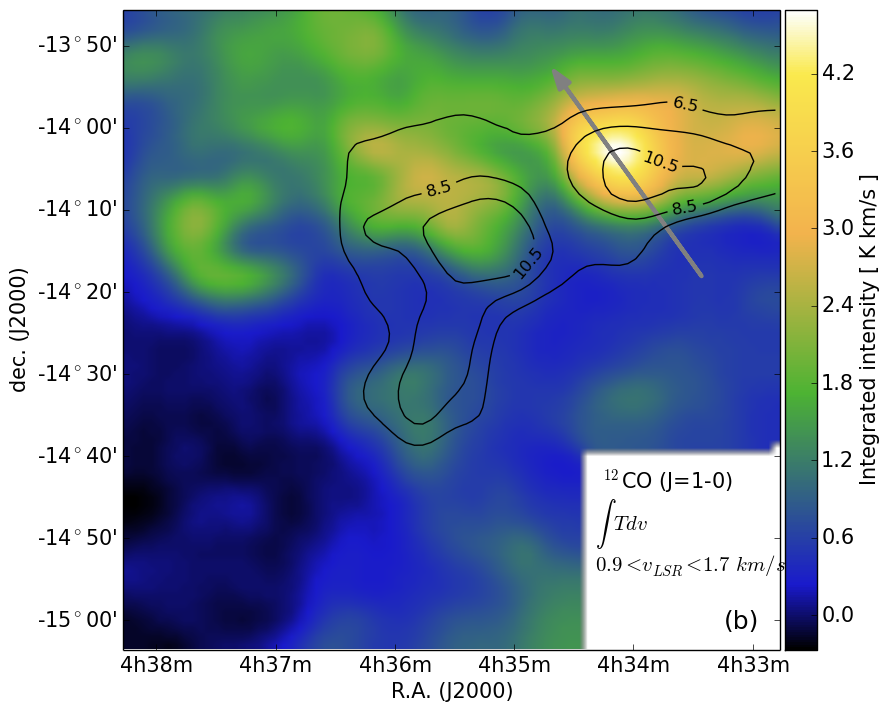}\\
\includegraphics[width=0.5\textwidth]{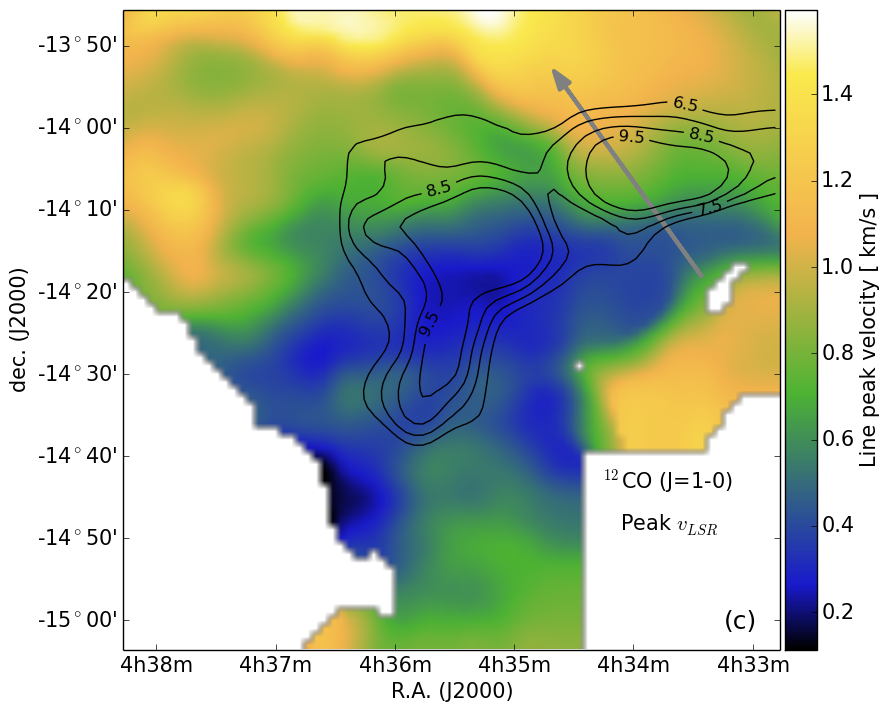}
\includegraphics[width=0.485\textwidth]{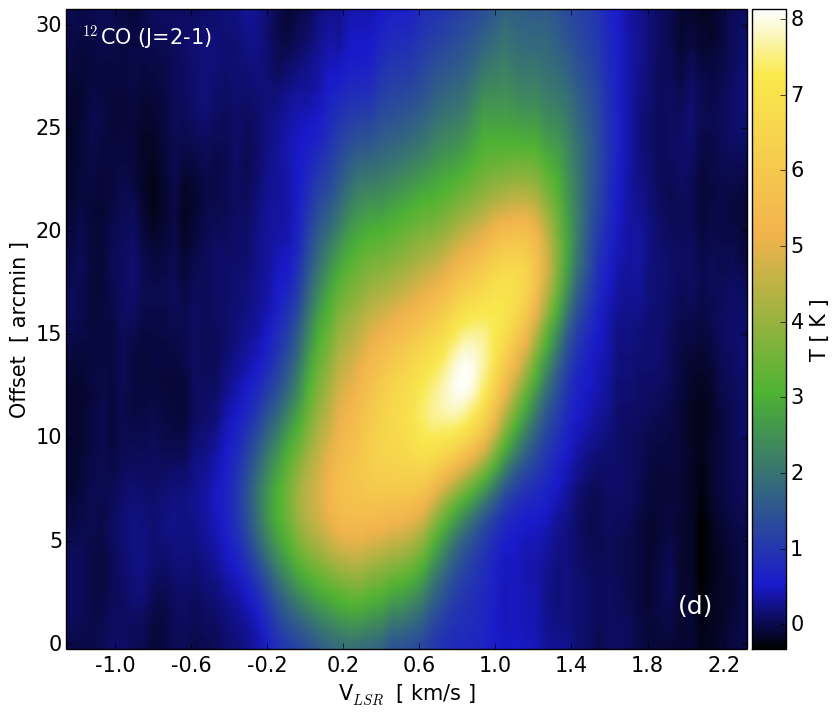}
\end{tabular}
\caption{$^{12}$CO data of L1642, based on the observations of~\citet{Russeil2003}. ($a$) Total integrated intensity, ($b$) intensity integrated between $0.9 < v_{\rm{LSR}} < 1.7$ km/s, ($c$) peak velocity obtained by a Gaussian fit of the $^{12}$CO$(1-0)$ line, ($d$) position-velocity diagram of $^{12}$CO$(2-1)$ along a cut through Clump A and a striation. The cut is shown with grey arrow in the other figures. $v_{\rm{LSR}}$ is the radial velocity relative to the local standard-of-rest (LSR) frame. Here, $T$ is the antenna temperature, i.e., it represents the intensity of the emission. The black contours show the total integrated intensity.}
\label{fig:CO}
\end{figure*}

We use the CO data presented and analysed by~\citet{Russeil2003} to compare with the $Herschel$ map and magnetic field orientations in terms of morphology and kinematics. The general morphology of the cloud revealed by the CO data is very similar to the $Herschel$ data shown in Fig.~\ref{fig:herschel} ($left$). It is dominated by the bright clumps A, B and C (Fig.~\ref{fig:CO}a). The Eastern parts of the main filamentary structures in the North (S1) and East (S2) of L1642 are also visible. The Western parts of these structures are overwhelmed by the emission of the main cloud and can be partly revealed when selecting the larger velocities (Fig.~\ref{fig:CO}b). However, the relatively low resolution of the CO data does not enable to resolve the striations in these structures, similarly to $Planck$ data.

The large scale kinematics was studied by~\citet{Russeil2003}. Their figure 
A.1 shows channel maps of $^{12}$CO$(1-0)$. One part of the cloud, including most 
of the central (B) and South (C) parts of the cloud, has velocities 
mainly between $V_{\rm LSR}=-0.1$ and 0.5 km s$^{-1}$, while another part, 
including mostly the West (A) and North regions of the cloud, has 
velocities between $V_{\rm LSR}=0.7$ and 1.5 km s$^{-1}$. The North and East 
filamentary structures (S1 and S2) are connected to this second, more red-shifted, structure. 
In Fig.~\ref{fig:CO}c we show the map of peak velocity obtained by a Gaussian fit of the $^{12}{\rm CO}(1-0)$ line.

Interestingly, at $V_{\rm LSR}>1$ km s$^{-1}$, the emission of all CO
transitions is dominated by the northern part of clump A, at the root of the 
striations of the northern structure S1 (Fig.~\ref{fig:CO}b). 
Fig.~\ref{fig:CO}d shows the position-velocity (PV) diagram in $^{12}$CO$(2-1)$
along a cut through Clump A (grey arrows in Fig.~\ref{fig:CO}). 
We find a velocity gradient of $\sim 1$
km s$^{-1}$ over the $\sim 15'$ (0.6 pc) width of Clump A (between offsets of 5$\arcmin$ 
and 20$\arcmin$). The velocity then tends to stabilize around 1.1 km s$^{-1}$ when penetrating the striation region, on the western edge of S1 (offsets $> 20\arcmin$).
The cut is shown also on the $Herschel$ map in Fig.~\ref{fig:herschel} ($right$), revealing that it goes through the Western head of Clump A, and parallel to the striations (S3) on the Northern part of Clump A. For comparison, we also made a slightly different cut, aligned with the striations of S3 and the B lines. This does not notably change the properties of the PV cut.

\section{Discussion} \label{sect:discussion}

\subsection{Connection between magnetic fields and cloud structure}

L1642 is one of the Orion outlying clouds located at the head of a long HI pillar structure. The cloud itself is elongated in the Equatorial E-W direction, forming a complex, partly cometary shaped structure. As shown in~\citet{Malinen2014}, there are three cores with YSO systems (two binary and one single) with approximately equal distances between the cores, along a straight line, and apparently in age order from the youngest one (B-2) in the west to the oldest (B-3) in the east. For a cloud with such low mass ($\sim72M_{\odot}$) and high latitude, L1642 has unusually high star formation efficiency, $\sim$7\% \citep{Malinen2014}.

The $Planck$ polarization data shown in Figs.~\ref{fig:herschel} ($right$) and~\ref{fig:herschelplanck} reveal an ordered magnetic field that pervades the cloud from NE to S. 
In the smallest scales of the $Herschel$ map, magnetic field is nearly perpendicular to the line connecting the YSOs, at least at the western part of the dense region B. 
The magnetic field orientation is also closely correlated with the diffuse material surrounding the cloud, as the low density striations are aligned with the magnetic field orientation.
There clearly is a complex interplay between the cloud structure and large scale magnetic fields revealed by $Planck$ polarization data. This suggests that the large scale magnetic field is closely linked to the formation and evolution of the cloud. However, it is an open question whether magnetic fields shape the cloud, or if the cloud material shapes the magnetic field, or if they both are simultaneously affected by some physical force.

Already~\citet{Taylor1982} predicted that a shearing flow of material is shaping the cloud. \citet{Malinen2014} discussed the connections between L1642 and its larger surroundings. To summarise, L1642 appears to be affected mostly by a flow coming from the NW/W, or the cloud itself is travelling to NW/W. Both the large scale Fig.~\ref{fig:planck_large} and the smaller scale Fig.~\ref{fig:herschelplanck} LIC maps show bending magnetic fields at the NW side of L1642, corresponding to the findings of~\citet{Malinen2014} that the cloud seems to be affected by a compressing force from the NW direction. The cloud morphology and the apparent compression that is shaping the cloud from the west suggest that star formation in L1642 may be externally triggered. The open question is, what is causing the force shaping the cloud and bending the magnetic field. \citet{Lee2009} studied star formation in a group of clouds (not including L1642) around the Orion-Eridanus Superbubble, and concluded that compression from the Superbubble can potentially have a long-range triggering influence causing star formation in affected clouds. L1642 is located between two large Bubbles, the Orion-Eridanus Superbubble and the Local Bubble~\citep[see][for a summary]{Alcala2008,Malinen2014}. It is possible that this extreme location is enough to cause the observed phenomena and be the trigger of star formation in this cloud.

This study clearly shows the connection between the large scale magnetic field and small scale cloud structure in L1642. However, the conditions affecting the cloud could mean that magnetic field has more effect here than in other regions. Also, the geometry of this object can be favourable to reveal such features, as it is quite isolated, in high latitude, and shows clear magnetic field orientation. Statistical studies comparing other regions combining $Planck$ and $Herschel$ data would be useful to determine how common these features are. Also, high-resolution polarization observations would reveal more details of the small-scale connections between density structures and magnetic fields.

\subsection{Dynamics of the cloud and its striations}

In Section~\ref{sect:CO_results} we reported the detection of a clear velocity gradient through 
Clump A with a continuous variation in velocity from the south edge of Clump A 
to the North-East, parallel to the close-by striation (S3). Since the angle between the striation and
the plane of the sky remains unknown, it is unclear whether the red-shifted gas 
is approaching Clump A or escaping from it. 

One possibility is that gas is photoevaporating from Clump A due to the 
surrounding interstellar radiation field. However, the dust temperature 
determined from $Herschel$ data in this part of the cloud, $\sim 15-16$ K,~\citep{Malinen2014,Montillaud2015} does not support the idea of UV-rich region, and no ionizing star is known in the vicinity of L1642.
On the other hand, the gas may well be much hotter than dust at the diffuse cloud surface. 
Assuming 100 K gas temperature, the sound speed is $\sim$1 km/s. In this simplified case, the gas would flow outwards approximately a few times 0.1 pc in timescales of a few 10$^5$ years, which is compatible with the extent of the striations and the typical lifetime of molecular clouds.

Alternatively, the gas north of Clump A could be infalling into Clump A along 
the striations and following the magnetic field lines. In~\citet{Russeil2003}, 
the total mass of the cloud was estimated to be $\sim 60 M_{\odot}$ based on the CO data. In~\citet{Malinen2014}, the total mass was estimated to be slightly higher, $\sim 72.1 M_{\odot}$, based on dust emission. If one assumes 
accretion onto a $10 M_{\odot}$ clump, the free-fall velocity at the distance of 
5$\arcmin$ should be $\sim 0.7$ km/s. This value would be in good agreement with 
the observed difference in velocity between the centre of Clump A and the 
striations if one assumes that the angle between the striations and the plane 
of the sky is not too small. However, the relevance of this assumption is not 
clear, and the observed velocity difference could be too large to be due solely 
to infall. 

Rotation of Clump A around an axis in almost East-West orientation and almost in the plane of 
the sky is another possible component in the dynamics of the cloud. If the gas 
were in a Keplerian orbit with a mass of $10 M_{\odot}$ and a radius of $0.2$pc, 
the velocity would be $\sim 0.5$ km/s. Therefore the combination of infall and 
rotation could explain the observed velocity gradient.

A last alternative we consider here is the collision between two gas streams. We 
mentioned in Sect.~\ref{sect:CO_results} that L1642 presents two structures with separated
velocity ranges, the most blue-shifted including the densest part of the cloud 
(Clump B) and the southern part (Clump C), while the red-shifted part includes 
the North and West regions (including Clump A), as well as the two filamentary structures S1 
(North, connected to Clump A) and S2 (East, connected to Clump B). This scenario 
offers the advantage of a complete and consistent view of the evolution, 
morphology and kinematics of the cloud. The cloud would originate from the 
material of the two flows jammed at their meeting point, and would naturally be 
cometary shaped. The red-shifted structure would correspond to one stream that 
is flowing around the North edge of the blue-shifted structure. The striations 
and the velocity gradient observed in Fig.~\ref{fig:CO} would originate from
the shearing between the two streams, and would be interpreted as a flow of gas 
away from Clump A. This would fit into the notion of~\citet{Malinen2014} that Clump A appears to be compressed from the SW, as the striations (S3) are seen only on the Northern side of the clump.
The alignment between the striations and the magnetic field 
lines would also originate from the shearing, assuming that magnetic field lines 
are frozen in the gas. Finally, this would also explain the unusually high star
formation efficiency of the cloud~\citep{Malinen2014}, considering its low mass and high latitude.

\section{Conclusions} \label{sect:conlusions}

We have compared the large scale magnetic field revealed by $Planck$ polarization maps and $Herschel$ submm dust emission maps in the high-latitude cloud L1642. We conclude that
\begin{enumerate}
\item there is a close connection between the cloud structure and the large scale magnetic field in L1642 and the surrounding region, suggesting that magnetic field is closely linked to the formation and evolution of the cloud
\item the connection between cloud structure and large scale magnetic field is seen even at the finest details of the cloud, most notably in the striations
\item the distribution of relative orientation between cloud structure and magnetic field lines in diffuse medium has one peak centered at $\sim$0$^{\circ}$, indicating that diffuse striations and B field are clearly aligned
\item dense medium presents a bimodal distribution of relative orientation centered at $\sim$0$^{\circ}$ and 90$^{\circ}$, but separate regions have different behaviours: the dense South part (C) is perpendicular to the B field, West (A) and North (D) exhibit structures aligned on the B field, and in the densest region (B), we cannot make any distinction on the orientation
\item there is a clear transition from aligned to perpendicular structures approximately at a column density of $N_{\rm{H}} = 1.6 \times 10^{21}\, {{\rm cm}}^{-2}$ (this equals $\sim 10^{21.7}\, {\rm{cm}}^{-2}$ when using the same convention for dust opacity as~\citet{Planck2015XXXV})
\item comparison to large scale $Planck$ polarization data at $\sim$10$'$ resolution is very useful even when looking at the finest structures in higher resolution data, e.g. $Herschel$ at $\sim$18.3$''$
\item CO rotational emission confirms that the striations are connected with the main clumps and likely to contain material either infalling to or flowing out of the clumps
\item Rolling Hough Transform, which was developed to extract linear features in large scale diffuse HI regions, is a very useful and practical method also when studying denser regions with more complex structure
\end{enumerate}

\section*{Acknowledgements}

We thank the referee for useful comments which improved the paper. We thank Kimmo Lehtinen and Delphine Russeil for providing us the CO data from their earlier studies.
MJ acknowledges the support of the Academy of Finland grants No. 285769 and 250741. S.E.C. was supported by a National Science Foundation Graduate Research Fellowship under grant No. DGE-11-44155.
The development of Planck has been supported by: ESA; CNES and CNRS/INSU-IN2P3-INP (France); ASI, CNR, and INAF (Italy); NASA and DoE (USA); STFC and UKSA (UK); CSIC, MICINN and JA (Spain); Tekes, AoF and CSC (Finland); DLR and MPG (Germany); CSA (Canada); DTU Space (Denmark); SER/SSO (Switzerland); RCN (Norway); SFI (Ireland); FCT/MCTES (Portugal); and The development of Planck has been supported by: ESA; CNES and CNRS/INSU-IN2P3-INP (France); ASI, CNR, and INAF (Italy); NASA and DoE (USA); STFC and UKSA (UK); CSIC, MICINN and JA (Spain); Tekes, AoF and CSC (Finland); DLR and MPG (Germany); CSA (Canada); DTU Space (Denmark); SER/SSO (Switzerland); RCN (Norway); SFI (Ireland); FCT/MCTES (Portugal); and PRACE (EU).





\bibliographystyle{mnras}
\bibliography{biblio_j_L1642_2}







\bsp	
\label{lastpage}
\end{document}